\documentclass{aastex}

\begin{document}

\title{Evolution of the pc-scale structure of PKS 1934-638 revisited: first science with the ASKAP and New Zealand telescopes}

\author{Tzioumis, A.K.}
\affil{CSIRO Astronomy and Space Science}
\affil{PO Box 76, Epping, NSW 1710, Australia}
\email{Tasso.Tzioumis@csiro.au}
\author{Tingay, S.J., Stansby, B.}
\affil{International Centre for Radio Astronomy Research, Curtin University of Technology}
\affil{GPO Box U1987, Perth Western Australia 6102, Australia}
\author{Reynolds, J.E., Phillips, C.J., Amy, S.W., Edwards, P.G., Bowen, M.A., Leach, M.R., Kesteven, M.J., Chung, Y., Stevens, J., Forsyth, A.R.}
\affil{CSIRO Astronomy and Space Science}
\affil{P.O. Box 76, Epping, NSW 1710, Australia}
\author{Gulyaev, S., Natusch, T.}
\affil{Institute for Radio Astronomy and Space Research, Auckland University of Technology}
\affil{Private Bag 92006, Auckland 1142, New Zealand}
\author{Macquart, J.-P., Reynolds, C., Wayth, R.B., Bignall, H.E., Hotan, A., Hotan, C., Godfrey, L.}
\affil{International Centre for Radio Astronomy Research, Curtin University of Technology}
\affil{GPO Box U1987, Perth Western Australia 6102, Australia}
\author{Ellingsen, S., Dickey, J., Blanchard, J., Lovell, J.E.J.}
\affil{School of Mathematics \& Physics}
\affil{Private Bag 37, University of Tasmania, Hobart, TAS 7001, Australia}

\begin{abstract}
We have studied the archetypal Gigahertz Peaked Spectrum  radio galaxy, PKS~1934$-$638, using the Australian Long Baseline Array, augmented with two new telescopes that greatly improve the angular resolution of the array. These VLBI observations represent the first scientific results from a new antenna in NZ and the first antenna of the Australian SKA Pathfinder (ASKAP).  A compact double radio source, PKS~1934$-$638, has been monitored over a period of 40 years, and the observation described here provides the latest datum, eight years after the previous observation, to aid in the study of the long-term evolution of the source structure.  We take advantage of these new long baselines to probe PKS~1934$-$638 at the relatively low frequency of 1.4 GHz, in order to examine the effects of optical depth on the structure of the radio source.  Optical depth effects, resulting in the observation of frequency dependent structure, may have previously been interpreted in terms of an expansion of the source as a function of time.  Expansion and frequency dependent effects are important to disentangle in order to estimate the age of PKS~1934$-$638.  We show that frequency dependent structure effects are likely to be important in PKS~1934$-$638 and present a simple two-dimensional synchrotron source model in which opacity effects due to synchrotron self-absorption are taken into account.  Evidence for expansion of the radio source over 40 years is therefore weak, with consequences for the estimated age of the radio source.
\end{abstract}


\section{Introduction}

PKS~1934$-$638 is the archetype of a class of radio galaxies known as GHz-Peaked Spectrum (GPS) radio galaxies.  The properties of this class are well described by \citet{odea98}.

PKS~1934$-$638 was first noted in terms of its strongly peaked radio spectrum at 1.4 GHz by \citet{bol63} and \citet{kel66}, indicating the likely existence of compact structure in the radio source \citep{sli63}.  Subsequent interferometric observations confirmed this \citep{gub71} and high angular resolution observations of PKS~1934$-$638 were made periodically over the next 20 years \citep{pre89,tzi5ghz,tzi02,kin94}, at a range of frequencies.  Most recently, \citet{oja04} presented dual-epoch 8.4 GHz VLBI observations of PKS~1934$-$638 as part of a combined analysis of $\sim$30 years of observations.  

\citet{oja04} estimated an expansion rate of 23$\pm10\mu$as/yr between the two compact components of radio emission that dominate the structure of PKS~1934$-$638, at a redshift of $z=0.183$ \citep{pen78} giving an apparent separation speed of $\sim 0.2\pm0.1c$, an order of magnitude higher than what is generally assumed for the hot spot advance speed of Cygnus A \citep{rea96}.  Throughout the history of VLBI observations of PKS~1934$-$638 it has been clear that the structure and flux density of the object evolve slowly, if at all.  This stability is consistent with the general properties of GPS radio galaxies \citep{odea98}. The flux density stability of PKS~1934$-$638, in particular, has led it to be used as the primary flux density calibrator for the Australia Telescope Compact Array at centimetre wavelengths. 

Because of the proposed physical nature of GPS sources, as the young progenitors of FR-I and FR-II radio galaxies \citep{odea98}, studies of the evolution of GPS sources are very important, in order to estimate their ages via measurement of expansion of the radio structure \citep{pol03}.  Long time series observations are required because of the slow rate of evolution in GPS radio galaxies and further, optical depth effects may complicate the interpretation of multi-epoch observations made at different frequencies.  Mechanisms such as free-free absorption and synchrotron self-absorption have been proposed as providing significant optical depth in GPS sources at radio wavelengths and these different mechanisms are potentially difficult to disentangle \citep{tin03}.  GPS radio spectra peak at GHz frequencies and the optical depths change rapidly over the frequency range 1 - 10 GHz.  One must therefore be wary when interpreting multi-epoch data obtained at different frequencies, as is the case for the historical VLBI data for PKS~1934$-$638.  Frequency dependent structure has been clearly seen in another GPS radio galaxy, CTD 93 \citep{sha99, nag06}.

\citet{oja04}, in their analysis of historical PKS~1934$-$638 VLBI data, raise the possibility that frequency dependent structure effects complicate their analysis.  In this paper we further examine the possibility of frequency dependent structure in PKS~1934$-$638 and find evidence that the apparent expansion of the source noted by \citet{oja04} can be explained as a consequence of these frequency dependent effects.  In section 2 we describe new VLBI observations at 1.4 GHz that use the Australian Long Baseline Array (LBA) augmented by two new radio telescopes.  In section 3 we discuss our new results within the context of the historical results (both those presented by Ojha et al. and other data we have extracted from the literature), in particular the possibility of frequency dependent structure in PKS~1934$-$638 and the impact on historical measurements of the expansion rate of PKS~1934$-$638.

\section{New VLBI observations and results}

VLBI observations of PKS~1934$-$638 were undertaken on 2010 April 29, using the array of radio telescopes listed in Table 1 and shown in Figure 1.  Two of these telescopes were used for the first time for scientific observations, the first ASKAP (Australian SKA Pathfinder) antenna in Western Australia \citep{joh07}, and Warkworth, a new facility of the Auckland University of Technology in New Zealand \citep{gul09}. The addition of ASKAP and Warkworth greatly increase the angular resolution (factor of $\sim$4) and $(u,v)$ coverage for observations of radio sources  by the Australian Long Baseline Array (LBA).  This is particularly useful at relatively low frequencies, where long baselines are required to obtain high angular resolution.

\begin{deluxetable}{lllcr}
\tablecaption{List of radio telescopes used in the VLBI observations.  ASKAP $=$ Australian Square Kilometre Array Pathfinder; ATCA $=$ Australia Telescope Compact Array.  Longitudes are listed as east of Greenwich.  SEFD $=$ System Equivalent Flux Density.}
\tablewidth{0pt}
\startdata
Telescope&Long. (deg)&Lat. (deg)&Diameter (m)&SEFD (Jy) \\ \hline
ASKAP&	116.64		&$-$26.69&				12&		6000 \\
Hobart&	147.44		&$-$42.80&				26&		600 \\
Parkes&	148.26		&$-$33.00&				64&		25 \\
Mopra&	149.07		&$-$31.30&		                  22&		350 \\
ATCA&		149.57	&$-$30.31&			5 $\times$ 22&	70 \\
Warkworth&	174.66	&$-$36.43&				12&		8000 \\ \hline
\enddata
\end{deluxetable}

\begin{figure}
\epsscale{0.5}
\plotone{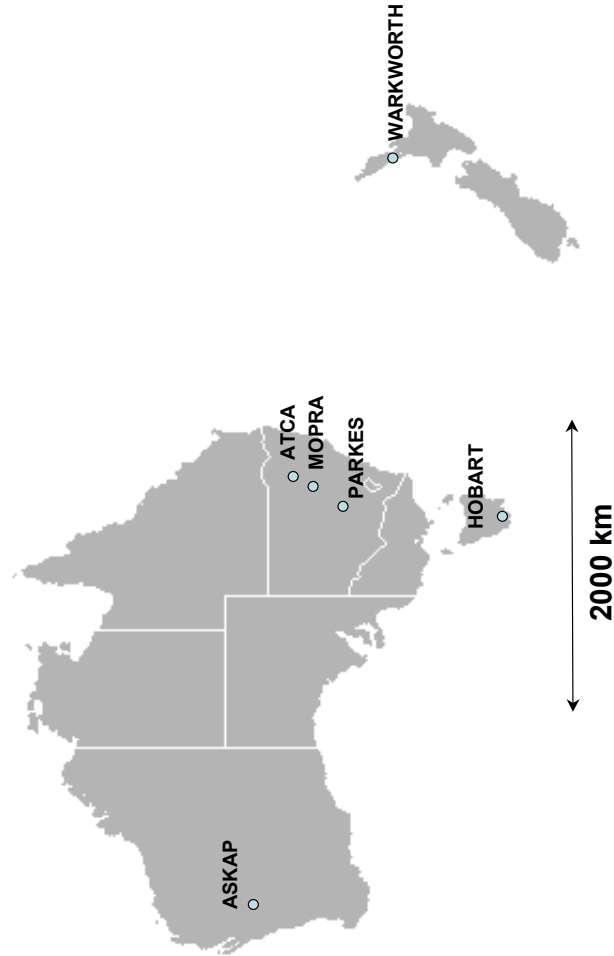}
\vspace{2cm}
\caption{Schematic geographical distribution of antennas used in the VLBI observations of PKS~1934$-$638}
\end{figure}

The VLBI observations were made by recording right and left circularly polarised signals at each antenna in the frequency range 1368 $-$ 1432 MHz (a 64 MHz bandwidth).  Data at the ATCA, Parkes, Mopra and Hobart antennas were recorded with the standard LBA systems \citep{phi09}.  The observation occurred over a period of six hours.

The data from the ASKAP and Warkworth antennas were obtained from a custom-made single-pixel feed and receiver system that delivered the band limited signals (downconverted to the frequency range 256 $-$ 320 MHz) at both polarisations to a custom-made digital system.  The digital system utilises a commodity Signatec sampler/digitiser PCI-based card (Signatec PX14400) to sample the band limited analog signals at the Nyquist-Shannon rate \citep{sha49} 
and digitise the signals with 14-bit precision.  These 14-bit samples are then recorded to a 32 TB RAID disk.  The Signatec card and RAID are hosted in a server class PC.  After recording, the 14-bit samples are converted to 2-bit precision, to decrease data storage and transport requirements, and to be more compatible with the 2-bit samples recorded with the standard LBA system.  A full description of the digital system will appear in \citet{sta10}.  A hydrogen maser clock was available at Warkworth and a rubidium clock was used at ASKAP, to drive frequency synthesisers for the local oscillator and the digital system sample clock.  Global Positioning System receivers at Warkworth and ASKAP were used to determine ($x,y,z$) positions for the antennas and generate 1PPS signals for the digital system.

Following the observations, data from most telescopes were transferred via fast network connections to Curtin University of Technology in Western Australia to be correlated using the DiFX software correlator \citep{del07}.  The data from the ASKAP telescope were transported to Perth by car, as fast network connections are still under construction. The data were correlated using an integration period of 2 seconds and 32 frequency channels per 64 MHz band.  All Stokes parameters were correlated.  PKS~1934$-$638 was detected with high signal to noise on all baselines in the array.  The resultant $(u,v)$ coverage for the observation is shown in Figure 2. All baselines above $6M\lambda$ are due to the new antennas at ASKAP and Warkworth, increasing the maximum baseline to ~$24M\lambda$ and thus the array resolution by a factor of ~4.

\begin{figure}
\epsscale{0.5}
\plotone{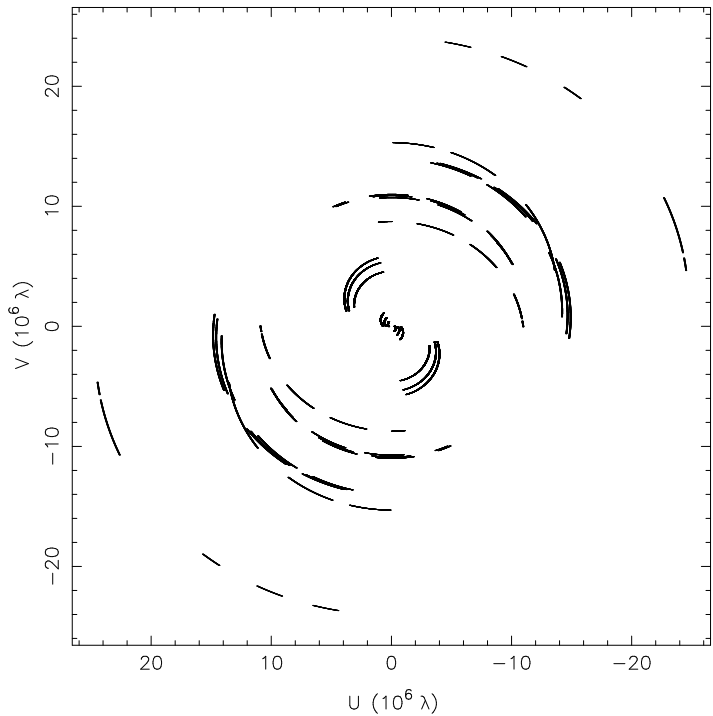}
\caption{$(u,v)$ coverage of the VLBI observation of PKS~1934$-$638. All baselines larger than $6M\lambda$ are contributed by the new antennas at ASKAP and Warkworth.}
\end{figure}

The correlated data were imported into AIPS for initial processing using standard techniques such as fringe-fitting.  The data were also amplitude calibrated using system temperature and gain values. For the new antennas at ASKAP and Warkworth these calibration parameters were measured at the telescopes by performing on-off pointings toward strong radio sources of known strength.  Once fringe-fitted and amplitude calibrated, the data were exported to DIFMAP \citep{she94} for excision of bad data, imaging and model-fitting.

Figure 3 shows the best representation of the Stokes I data in the image plane, obtained by model-fitting the visibility data, using the model specified in Table 2, determined using the modelfit task in DIFMAP.  A uniform weighting of the $(u,v)$ data was used in the transformation to the image plane.  Phase self-calibration was used in early iterations of model-fitting, with amplitude self-calibration used in the final iterations of model-fitting.  Amplitude self-calibration led to changes in individual antenna gains of the order 15\%, relative to the {\it a priori} amplitude calibration.

\begin{figure}
\plotone{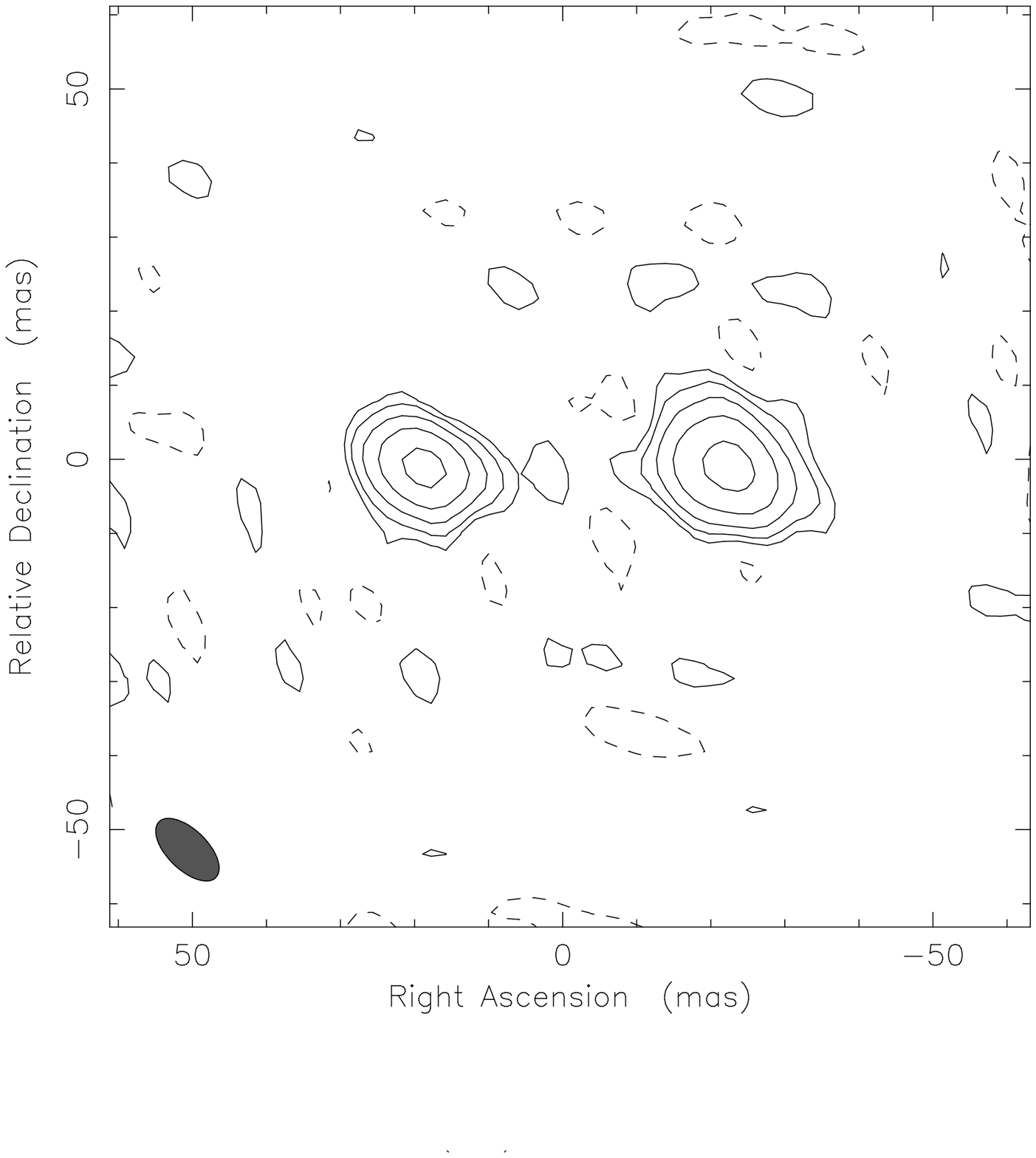}
\caption{Stokes I VLBI image of PKS~1934$-$638 at 1.4 GHz.  Contour levels are $-$4 ,4, 8, 16, 32 and 64\% of the peak surface brightness of 2.4 Jy/beam.  The restoring beam is 10.8 $\times$ 5.4 mas at a position angle of 45.7${^\circ}$.  Note, the phase centre of the image has been shifted relative to the positions of the components indicated in Table 2 by (-20 mas, 0 mas), for convenience of display.}
\end{figure}

\begin{deluxetable}{crrccr}
\tablecaption{Model for the structure of PKS~1934$-$638 at 1.4 GHz.  $S =$ flux density of model component; $R =$ distance of model component from phase centre of image; $\theta =$ position angle of model component centroid from the phase centre of image; $a =$ major axis length of model component; $r =$ minor axis to major axis ratio; $\phi =$ position angle of model component major axis}
\tablewidth{0pt}
\startdata
$S$ (Jy) &$R$ (mas)  &$\theta$ (deg)  &$a$ (mas)  &$r$   &$\phi$ (deg) \\ \hline
5.6  &   0.0  &    0.0    &  8.7  &  0.9  & 6.4 \\
3.7  &   41.0 &    $-$90.7  &   8.2 &    0.0 &  $-$62.5 \\ \hline
\enddata
\end{deluxetable}

The structure seen in Figure 3 agrees very well with all previous published images of PKS~1934$-$638 in this angular resolution regime with synthesised beams of $\approx$ 5-10 mas \citep{tzi89,kin94,tzi5ghz,tzi02,oja04,vsop}.  The two components have been interpreted as the terminal hot spots from oppositely directed jets emerging from a black hole accretion disk system, a smaller and younger version of an FR-I or FR-II type radio galaxy, into which GPS radio galaxies are postulated to evolve.  This is the general interpretation of these double component structures, which are typical of GPS radio galaxies \citep{odea98}.

We derived errors on the main parameter of interest in this analysis, the separation between the two components in the image, in order to compare our results to historical results.  We did this by keeping one model component fixed at the phase centre of the image and forcing variation of the position of the second component, then assessing the degradation of the fit of data to model as a function of this variation.  This technique for estimating errors has been used extensively previously \citep{tin98}.  In the case of PKS~1934$-$648, which is an almost equal strength double at this angular resolution, the visibility amplitudes and phases vary strongly, allowing this technique to tightly constrain the error bars on the separation, especially using the long and sensitive baselines from the east coast Australian antennas to ASKAP and Warkworth.  We estimated that the separation of the two components at 1.4 GHz is 41.0 mas $\pm$ 0.6 mas (3$\sigma$).

\section{Discussion}

To place our new measurement of the separation between the two compact components in PKS~1934$-$638 within the context of the historical measurements, we reproduce in Table 3 the data compiled by \citet{oja04}, with the addition of our new datum at 1.4 GHz.  Also listed in Table 3 are the results of other historical observations that were not included in the analysis of \citet{oja04}, namely a 5 GHz observation from the Southern Hemisphere VLBI Experiment (a precursor of the Australian LBA) \citep{tzi5ghz} and a 1.6 GHz observation from the VLBI Space Observatory Programme \citep{vsop}. In addition, the parameter measurements for the images of 1998 and 1991 \citep{kin94, tzi5ghz, tzi02} have been repeated using the original published images, and new estimates of separation and position angle are reported, together with conservative error estimates. Table 3 now includes all historical VLBI observations of PKS~1934$-$638, with all parameters reported in a consistent way.

\begin{deluxetable}{cllcc}
\tablewidth{0pt}
\tablecaption{Measured component separations in PKS~1934$-$638.  References are: 1=\citet{gub71}; 2=\citet{tzi89}; 3=\citet{kin94}; 4=\citet{tzi02}; 5=\citet{tzi5ghz}; 6=\citet{vsop}; 7=\citet{oja04}; 8=This work.}
\startdata
Epoch  &Separation& Position Angle &       Freq (GHz) & references \\ \hline
1970.8 &41.9$\pm$0.2 &$-$90$\pm$0.1  & 2.3 & 1 \\
1982.3 &42.0$\pm$0.2 &$-$90.5$\pm$0.1 & 2.3 & 2 \\
1988.9 &41.4$\pm$0.5 &$-$92$\pm$0.5  & 2.3 & 3,4,5 \\
1991.9 &42.2$\pm$0.5 &$-$92$\pm$0.5  & 8.4 & 3,4,5 \\
1992.9 &41.9$\pm$0.5 &$-$92$\pm$0.5  &4.85 & 5 \\
1999.3 &41.4$\pm$0.5 &$-$92$\pm$0.5 &1.65 & 6 \\
2002.5 &42.7$\pm$0.4 &$-$92$\pm$0.5  &8.4& 7 \\
2002.9 &42.6$\pm$0.3 &$-$92$\pm$0.4  &8.4& 7 \\
2010.3 &41.0$\pm$0.6 &$-$91$\pm$1.0  & 1.4& 8 \\ \hline
\enddata
\end{deluxetable}

We reproduce Figure 2 of \citet{oja04} below in Figure 4, including the new datum and the additional historical data. Also plotted are lines showing the expansion rate of  23$\pm10\mu$as/yr between the two compact components of radio emission in PKS~1934$-$638, as estimated in \citet{oja04}.
As can be seen from Table 3 and Figure 4, the new measurement of separation at 1.4 GHz lies significantly below the extrapolation of the previously measured values of separation at 2.3 and 8.4 GHz.  The same can be said of the 1.6 GHz VSOP datum.  However, if the separations are plotted as a function of observing frequency, as shown in Figure 5, a consistent trend of separation with frequency is revealed, with larger separations measured at higher frequencies and smaller separations measured at lower frequencies.

\begin{figure}
\plotone{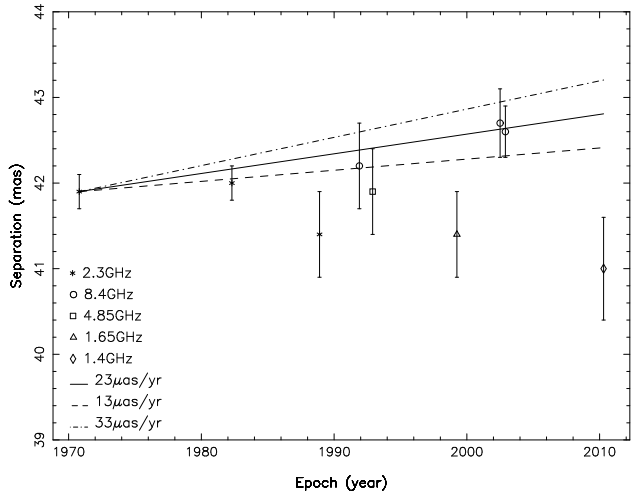}
\caption{Separation between the two components in PKS~1934$-$638, as a function of time. The solid lines show the predictions of the expansion rate estimated in \citet{oja04}.}
\end{figure}

\begin{figure}
\epsscale{1}
\plotone{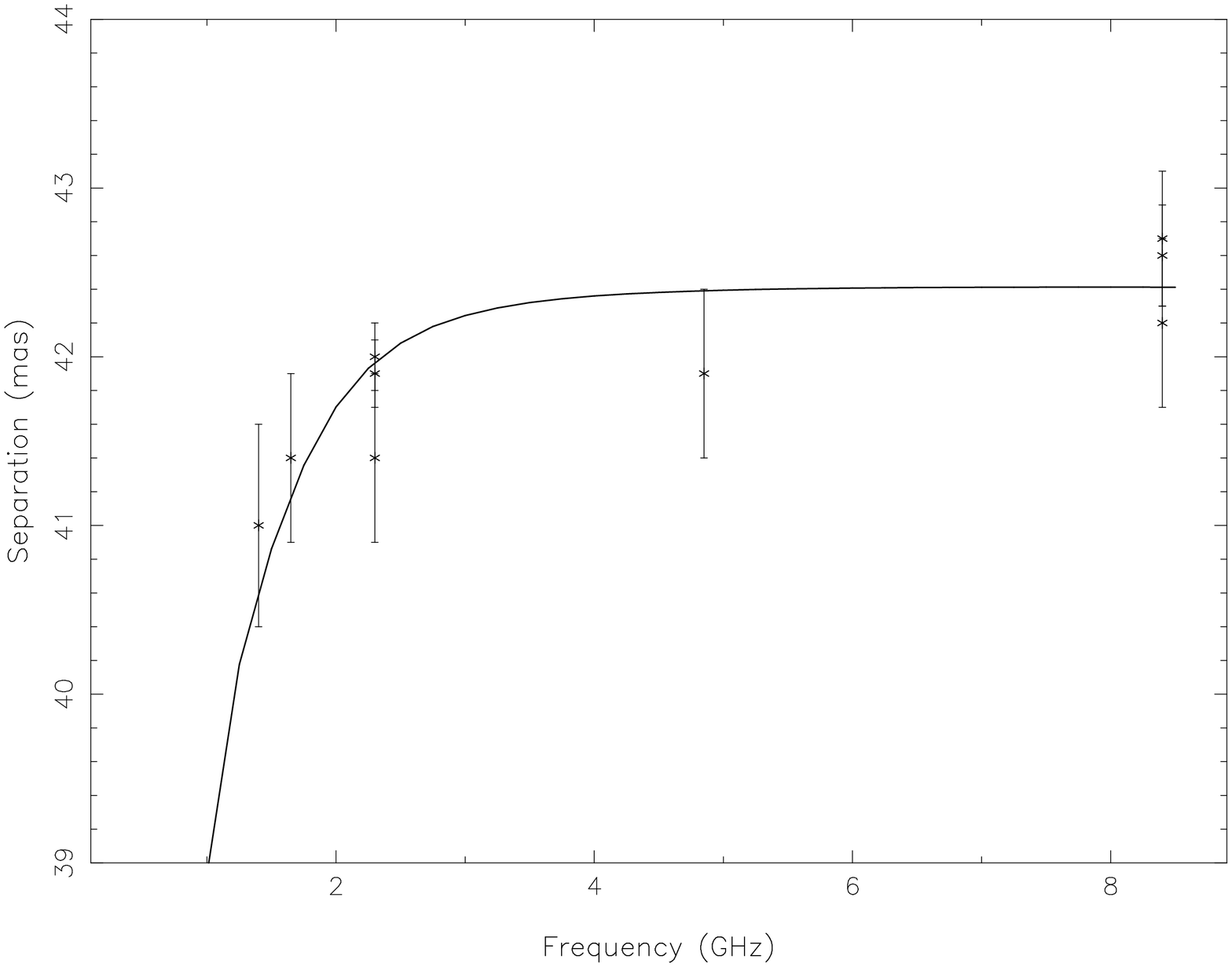}
\caption{Separation between the two components in PKS~1934$-$638, as a function of observing frequency. The results of the model described in the text  is plotted for a viewing angle $\theta=0$ (solid line), to explain the variation in separation as a function of frequency. }
\end{figure}

We suggest that this is a signature of frequency dependent structure in PKS~1934$-$638 and is caused by optical depth effects.  To investigate the origin of the apparent increase of lobe separation with frequency, we explored a two-dimensional synchrotron source model which takes into account opacity effects associated with synchrotron self-absorption.  We modelled the electron density profiles of the two lobes in terms of a ``cut-off'' gaussian profile as a means of representing the discontinuity associated with the shock front at the jet working surface and the corresponding backflow from the shock region. In the plane of the jet axis the lobe edges are located at $\pm x_0$, measured relative to the core at $x=0$, so that the electron density profiles of the two lobes are:
   
\begin{equation}
\rho(x,z) =  \left\{ \begin{array}{ll} \rho_0 \exp \left[ - \frac{(|x| -  
x_0)^2 + z^2}{r_0^2} \right],  &  |x| < x_0, \\
0, & \hbox{otherwise}, \\
\end{array} \right. \label{rhoProfile} \\
\end{equation} 

A constant magnetic field was assumed, and a spectral index of $-3.7$ for the electron energy distribution was chosen  to match that implied by the  $\nu^{-1.35}$ spectrum of the optically thin regime at frequencies above $2\,$GHz.   The spectral index was assumed constant across the extent of the source; the model does not take into account any spectral steepening due to synchrotron cooling in the regions removed from the shock front.  Correction for an arbitrary viewing angle, $\theta$, was made by effecting a rotation of the co-ordinate axes $(x',z') \rightarrow (x \cos \theta + z \sin \theta, -x \sin \theta + z \cos \theta)$, and integrating along the $z'$ axis the equations of radiative transfer \citep{saz}:

\begin{eqnarray}
{d\over d z'}\pmatrix{I\cr Q \cr
U}= \pmatrix{\alpha_I\cr \alpha_Q\cr 0}
+\pmatrix{-\mu_I&-\mu_Q&0\cr -\mu_Q&-\mu_I&-\rho_V\cr
0&\rho_V&-\mu_I}\pmatrix{I\cr Q\cr U}. \label{IQUmatrix}
\end{eqnarray}
Here, the emission and absorption coefficients are $\alpha_{I,Q}$ and $\mu_{I,Q}$ respectively  \citep{mel91}.  We neglect the contribution of Faraday rotation associated with any ambient thermal plasma and set $\rho_V=0$.  

The lobe separation was computed by calculating the separation between the intensity-weighted means of the emission around each peak in the emission profile.   The value of $x_0$ was fixed by requiring that, for a given viewing angle, the radiative transfer code reproduced the observed separation between the lobes at high frequency, where the lobe separation asymptotes to the true lobe separation. 

This simple model explains the data to the satisfaction of the error bars.We find that models within $\sim 20^\circ$ of the plane of the sky account well for the observed frequency dependence of the lobe separation. Thus, we can demonstrate that our suggestion of frequency-dependent structure has a reasonable basis in physics.  A corollary of this suggestion is that the expansion rate calculated by \citet{oja04} should be viewed with considerable caution, until more VLBI measurements can be made at a single frequency, to minimise frequency dependent structure effects.  We note that the sequence of 2.3 GHz observations listed in Table 3, and shown in Figure 4, by themselves do not constitute evidence for a significant expansion rate.  Likewise, the sequence of 8.4 GHz observations, taken by themselves, do not show evidence for significant expansion.  Given the evidence for frequency dependent source structure, we consider the evidence for source expansion to be weak.  A further consequence of the frequency dependent source structure is that estimates of the radio source age derived from the apparent expansion must also be considered with caution.

\section{Conclusions}

We have obtained new high angular resolution observations of the archetype GPS radio galaxy, PKS~1934$-$638, by using the Australian Long Baseline Array enhanced with two new radio telescopes, the first antenna of the Australian SKA Pathfinder in Western Australia and a new 12 m antenna at Warkworth, New Zealand.  The addition of these two new antennas have greatly improved the angular resolution obtainable at the low frequency end of the the LBA's operating range, 1.4 GHz.  These new capabilities have been used to examine the frequency dependent structure of PKS~1934$-$638 on the parsec scale, within the context of historical VLBI observations of the source at various frequencies.  
We show that frequency dependent structure effects are important in PKS~1934$-$638 and present a simple two-dimensional synchrotron source model in which opacity effects due to synchrotron self-absorption are taken into account. We find that previous estimates of the expansion rate of the radio source must be viewed with caution,  and therefore that calculations of the source age and limits of the onset of jet production must also be considered with care.

\acknowledgments
The International Centre for Radio Astronomy Research is a Joint Venture between Curtin University of Technology and The University of Western Australia, funded by the State Government of Western Australia and the Joint Venture Partners.  The Australian Long Baseline Array is part of the Australia Telescope which is funded by the Commonwealth of Australia for operation as a National Facility managed by CSIRO.  Mr Bruce Stansby is supported by a PhD scholarship provided by Curtin University of Technology.  Prof. Steven Tingay is a Western Australian Premier's Fellow.  The authors acknowledge the role of the Australia New Zealand SKA Coordination Committee in helping to foster and coordinate radio astronomy developments between Australia and New Zealand. This scientific work uses data obtained from the Murchison Radio-astronomy Observatory. We acknowledge the Wajarri Yamatji people as the traditional owners of the Observatory site.

\end{document}